\documentclass[pre,twocolumn,nofootinbib,twoside,showpacs,superscriptaddress,tightenlines,floatfix,aps]{revtex4-2}
\usepackage[T1]{fontenc}
\usepackage[utf8x]{inputenc}
\usepackage[T1]{fontenc}
\usepackage[normalem]{ulem}
\usepackage{color}
\usepackage{bm}
\usepackage{tikz}
\usepackage{mathtools}
\usepackage{dsfont}
\usepackage{nicefrac}
\usepackage{hyperref}
\hypersetup{
    allcolors=blue,
    colorlinks=true,
    bookmarks=true,
    pdfpagemode=FullScreen,
}
\usepackage{ulem}
\usepackage{tikz}
\usepackage[compat=1.1.0]{tikz-feynman}
\usepackage{graphicx} 
\usepackage{bm}
\usepackage{physics}
\usepackage{txfonts}
\usepackage{mathrsfs}
\usepackage{amsmath}
\usepackage{feynmf}
\usepackage{comment}
\usepackage{xcolor}
\usepackage{upgreek}
\usepackage{hyperref}
\hypersetup{
    allcolors=blue,
    colorlinks=true,
    bookmarks=true,
    pdfpagemode=FullScreen,
}

\begin{document}
\title{Supersymmetry Without Time-Reversal Invariance in Model A: A FRG perspective}

\author{Sankarshan Sahu}
\affiliation{
Laboratoire de Physique Th\'eorique de la Mati\`ere Condens\'ee, UPMC,
CNRS UMR 7600, Sorbonne Universit\'e, 4, place Jussieu, 75252 Paris Cedex 05, France
}
\author{Bertrand Delamotte}
\affiliation{
Laboratoire de Physique Th\'eorique de la Mati\`ere Condens\'ee, UPMC,
CNRS UMR 7600, Sorbonne Universit\'e, 4, place Jussieu, 75252 Paris Cedex 05, France
}
\author{Adam Ran\c con}
 \affiliation{
Univ.\ Lille, CNRS, UMR 8523 -- PhLAM -- Laboratoire de
Physique des Lasers Atomes et Mol\'ecules, F-59000 Lille, France
}
\affiliation{Institut Universitaire de France}
\author{Matthieu Tissier}
\affiliation{
Laboratoire de Physique Th\'eorique de la Mati\`ere Condens\'ee, UPMC,
CNRS UMR 7600, Sorbonne Universit\'e, 4, place Jussieu, 75252 Paris Cedex 05, France
}
\date{\today}
\begin{abstract}
We show that, contrary to common belief, supersymmetry alone is not sufficient in Model A dynamics to ensure relaxation toward a stationary state satisfying time-reversal invariance (TRI). An additional condition on top of supersymmetry is required for TRI, which we analyze in detail. We explicitly construct a model that is supersymmetric but violates TRI, and argue that, at least perturbatively, TRI nevertheless emerges as an effective large-scale symmetry. Using the functional renormalization group (FRG), we further show that the dynamical effective action, $\Gamma[\varphi,\tilde\varphi]$, contains the derivative of the equilibrium effective action, $\Gamma^{\mathrm{eq}}[\varphi]$, whose renormalization-group flow is identical to that of the equilibrium theory order by order in the derivative expansion. Finally, extending the same line of reasoning, we show that the probability distribution of the total magnetization in the Ising model can be recovered within the Model A framework.
\end{abstract}

\maketitle

\section{Introduction}
    All $N$-body statistical systems satisfying the detailed balance relation have time-reversible dynamics in the steady state. This property, which implies that no steady-state currents exist, distinguishes them from other systems that reach a steady state at large time. 
    
    In a non-trivial way, and for Langevin dynamics, these systems can also be described at mesoscopic scales by field theories -- the Martin-Siggia-Rose-de Dominicis-Jannssen (MSRDJ) \cite{PhysRevA.8.423, PhysRevB.18.4913, Janssen:1976qag} method -- satisfying supersymmetry. This symmetry is more mysterious because it involves auxiliary, non-physical, Grassmann fields, which are only brought into play to take into account a Jacobian in the construction of the field theory. These fields can be completely eliminated from the MSRDJ field theory using It$\bar{\rm o}$'s formulation of stochastic calculus, but supersymmetry is then hidden, much like the gauge symmetry of electromagnetism once the Coulomb gauge is chosen. 
    
    The popular belief is nevertheless that supersymmetry expresses nothing more than time reversal symmetry  in the MSRDJ field theory involving Grassmann fields, when no particular stochastic calculus choice is made. There would thus be a trade-off between the simplicity of It$\bar{\rm o}$'s calculus, which does not involve any Grassmannian fields -- and which generalizes to non-thermal stationary systems -- and the abandonment of explicit covariance under supersymmetry.

In what follows, we consider $\mathbb Z_2$-symmetric scalar models whose dynamics are governed by a Langevin equation with Gaussian noise. Although we restrict the discussion to this class of systems, most of our arguments can be extended to models that reach thermodynamic equilibrium at long times. This article has four main goals.

(i) We first show that supersymmetry alone does not guarantee that the equilibrium dynamics is time-reversible.

(ii) Using It$\bar{\rm o}$'s prescription, we then prove that, order by order in the derivative expansion, the renormalization-group (RG) flow of $\Gamma^{(0,1)}[\varphi(x,t)=\varphi(x),\tilde\varphi(x,t)=0]$ is closed, where $\Gamma$ denotes the effective action and $\tilde\varphi(x,t)$ the response field. In other words, this flow is independent of $\Gamma^{(0,n)}[\varphi(x,t),\tilde\varphi(x,t)=0]$ with $n>1$.

(iii) We further show that this flow coincides with the RG flow of the (derivative of the) equilibrium effective action --  or Gibbs free energy -- $\Gamma^{\rm\, eq}[\varphi]$.

(iv) Finally, we demonstrate that the RG flow of the equilibrium probability distribution of the order parameter -- $L^{-d}\int_x\, \phi(x)$
 -- can be obtained from model A in the long-time limit.

Point (i) is nontrivial since it implies the existence of models that satisfy supersymmetry while breaking TRI. From an RG viewpoint, this observation is intriguing. It indicates that, within the space of $\mathbb Z_2$-symmetric scalar models, the supersymmetric sector forms a subspace that is stable under the RG flow but does not coincide with the set of models satisfying a time-reversal dynamics {\it \`a la} model A. The latter defines a smaller RG-stable subspace within the supersymmetric one.

We show that for $\mathbb Z_2$-symmetric scalar systems, the operators that preserve supersymmetry while breaking TRS are highly irrelevant in the RG sense. This strongly suggests that, at criticality, the RG flow is attracted to the usual Wilson--Fisher fixed point, where TRS is effectively restored at long distances and long times.

Points (ii) and (iii) are not new \cite{general-framework, Roth:2024rbi}. However, we provide a proof that is particularly well suited to the functional renormalization group (FRG), and in particular to its most widely used approximation scheme, the derivative expansion. \footnote{parallel with translation invariance}

At first sight, point (iv) may appear as a straightforward consequence of points (ii) and (iii), and in a sense it is. However, we are not aware of any alternative derivation of this result. Indeed, at equilibrium, one can derive from first principles the relation between the probability distribution function (PDF) of the order parameter and (a slightly modified version of) the Gibbs free energy, that is, of the effective action. By contrast, no such relation has been established for model A. To the best of our knowledge, the argument given in point (iv), although indirect, provides the only derivation of this result to date. 

\section{Model and supersymmetry}

We consider Model A dynamics, i.e., a dynamical $\phi^4$ theory that relaxes toward thermal equilibrium at long times. The dynamics is governed by the Langevin equation
\begin{equation}
    \partial_t \phi(x,t) = -\frac{\delta H[\phi]}{\delta \phi(x,t)} + \zeta(x,t),
    \label{langevin}
\end{equation}
where $\zeta$ is a Gaussian noise and $H$ is a general Hamiltonian that can for instance be the standard $\phi^4$ Hamiltonian:
\begin{equation}
    H[\phi] = \int_{x} \left[ \frac{1}{2}(\nabla \phi)^2 + V(\phi) \right],
    \qquad 
    V(\phi) = \frac{r}{2}\phi^2 + \frac{u}{4!}\phi^4 .
    \label{hamiltonian}
\end{equation}
The noise has zero mean and variance
\begin{equation}
    \langle \zeta(x,t)\zeta(x',t') \rangle
    = 2\,\delta(t-t')\,\delta^{d}(x-x').
    \label{noise}
\end{equation}

Using the Martin--Siggia--Rose--de~Dominicis--Janssen (MSRDJ) formalism, the generating functional of correlation and response functions can be derived from the Langevin equation~\eqref{langevin}. At this stage two convenient choices can be made. One may adopt It$\bar{\rm o}$ discretization, in which case the Jacobian appearing in the MSRDJ construction is equal to unity and can therefore be ignored. Alternatively, the Jacobian can be expressed in terms of Grassmann fields; the resulting supersymmetric formulation allows one to treat all fields simultaneously through the introduction of a superfield that collects them.

\subsection{MSRDJ formulation}
\label{msdrdj}
Using It$\bar{\rm o}$ calculus, the generating functional reads
\begin{equation}
    Z[j,\tilde j] =
    \int D\phi\, D[i\tilde\phi]\,
    \exp\left(
        -S[\phi,\tilde\phi]
        + \int_{x,t} (j\phi + \tilde j\,\tilde\phi)
    \right),
    \label{Z-nonequilibrium}
\end{equation}
with action
\begin{equation}
    S[\phi,\tilde\phi]
    =
    \int_{x,t}
    \left[
        \tilde\phi
        \left(
            \partial_t \phi
            +\frac{\delta H[\phi]}{\delta\phi}
        \right)
        - \tilde\phi^2
    \right].
    \label{action-Ito}
\end{equation}

Within this formalism, TRI corresponds to an invariance of the action under the transformation
\begin{align}\label{trs}
   T(t)=-t,
\quad  T(\phi(t))=\phi(-t)\nonumber\\
\quad
T(\tilde\phi(t))=\tilde\phi(-t) -  \dot\phi(-t) 
\end{align}
where the $x$-dependence of the fields has not been mentioned to alleviate the notation.

  Let us notice that the action $S$ is proportional to $\tilde\phi$ which, in It$\bar{\rm o}$, is a consequence of causality  and it can be shown \cite{general-framework} that this property holds for all 1PI correlation functions and thus for the effective action.
  
\subsection{Supersymmetric formulation}

In the supersymmetric formulation the generating functional becomes
\begin{equation}
    Z[J] =
    \int D\Phi\,
    \exp\left(
        -{\cal S}[\Phi]
        + \int_{\varpi} J \Phi
    \right),
\end{equation}
where $\Phi$ is the superfield
\begin{align}
& \Phi(t,x,\theta,\bar\theta)\nonumber\\
& \equiv \Phi(\varpi)
=
\phi(t,x)
+ \eta(t,x)\bar\theta
+ \theta\,\bar\eta(t,x)
+ \theta\bar\theta\,\tilde\phi(t,x),
\label{superfield}
\end{align}
with $\varpi = (t,x,\theta,\bar\theta)$ and where $\theta$ and $\bar\theta$ are anticommuting Grassmann variables.

The supersource $J$ collects the sources $j$ and $\tilde j$ together with the Grassmann sources $\gamma$ and $\bar\gamma$,
\[
J
=
\tilde j
+ \gamma \bar\theta
+ \theta\bar\gamma
+ \theta\bar\theta\, j .
\]

The action reads
\begin{equation}
{\cal S}[\Phi]
=
\int_{\varpi}
\left(
    \bar D \Phi \, D \Phi
    + H(\Phi)
\right),
\label{actionsusy}
\end{equation}
where the covariant derivatives are defined as
\begin{equation}
D = \partial_\theta - \bar\theta \partial_t,
\qquad
\bar D = \partial_{\bar\theta},
\qquad
\{D,\bar D\} = -\partial_t 
\label{cov-der}
\end{equation}
and
\begin{equation}
    \int_{\varpi}=\int d^dx\, dt\, d\bar\theta\, d\theta.
\end{equation}
The action ${\cal S}$ is invariant under supersymmetry transformations generated by
\begin{equation}
Q = \partial_\theta,
\qquad
\bar Q = \partial_{\bar\theta} + \theta \partial_t,
\qquad
\{Q,\bar Q\} = \partial_t .
\label{generators}
\end{equation}

\section{The functional renormalization group}

Several arguments presented below rely on the functional renormalization group (FRG). 
We briefly introduce here the basic elements of this method and refer to 
\cite{general-framework} for a comprehensive presentation of the formalism in the 
context of dynamical systems, both in the It$\bar{\rm o}$ and supersymmetric formulations.

At equilibrium, Wilson's implementation of the renormalization group consists in 
integrating out fluctuations in the partition function progressively, momentum 
shell by momentum shell. To this end, a momentum scale $k$ is introduced such that 
fluctuations with momenta larger than $k$ (rapid modes) are integrated out, while 
fluctuations with momenta smaller than $k$ (slow modes) are frozen. 

In the modern implementation of Wilson's RG, this procedure is realized by deforming 
the system through the addition of a quadratic term to the Hamiltonian:
\begin{equation}
    H[\phi] \to H[\phi] + \Delta H_k[\phi],
    \ \ 
    \Delta H_k[\phi] =
    \frac{1}{2}\int_q \phi(q)\, R_k(q)\, \phi(-q),
    \label{modified-H}
\end{equation}
in such a way that (i) for $k\simeq\Lambda$, where $\Lambda$ denotes the ultraviolet 
(UV) cutoff, i.e. the short-distance cutoff, all fluctuations are effectively 
frozen, and (ii) when $k=0$ all fluctuations are integrated out and the partition 
function of the original model is recovered. 

This is achieved by choosing a regulator function $R_k(q)$ satisfying these two 
requirements. A convenient choice is for instance
\begin{equation}
    R_k(q) = (k^2 - q^2)\,\theta(k^2 - q^2),
    \label{regulator-theta}
\end{equation}
although the precise shape of $R_k(q)$ does not play a role in what follows.  The form  \eqref{regulator-theta} effectively provides a $k$-dependent mass for the 
modes $\phi(q)$,  which almost freezes all fluctuations when $k$ is large  and vanishes when $k=0$ such that all fluctuations are integrated as in the original model.

The partition function of the deformed (or regularized) model reads
\begin{equation}
    Z_k[j] =
    \int D\phi\,
    e^{-H[\phi] - \Delta H_k[\phi] + \int_x j \phi}.
    \label{partition-function-k}
\end{equation}

It is now well understood that the RG flow, i.e. the evolution of physical 
quantities as $k$ is varied, is best controlled when approximations  are implemented at the level of the Legendre transform of $\log Z_k$, rather than 
directly on $\log Z_k$ itself. We therefore introduce the Gibbs free energy $\Gamma_k^{\rm Leg.}[\langle\phi\rangle]$ which is the Legendre transform
 of $\log Z_k$ defined by
\begin{equation}
    \Gamma_k^{\rm Leg.}[\varphi] + \log Z_k[j]
    =
    \int_x j(x)\,\varphi(x),
\end{equation}
where $\varphi(x)=\langle \phi(x) \rangle$. One then defines the modified 
Legendre transform
\begin{equation}
    \Gamma_k[\varphi]
    =
    \Gamma_k^{\rm Leg.}[\varphi]
    - \frac{1}{2}\int_q \varphi(q)\,R_k(q)\,\varphi(-q),
    \label{gamma-k-def}
\end{equation}
which is the quantity whose RG flow will be studied.

By construction, $\Gamma_{k=\Lambda}\simeq H$, since $R_\Lambda(q)$ is very large for 
all values of $q$, while $\Gamma_{k=0}=\Gamma$ corresponds to the Gibbs free 
energy of the model. The evolution of $\Gamma_k[\varphi]$ with the scale $k$ 
follows from Eqs.~\eqref{modified-H}, \eqref{partition-function-k}, and 
\eqref{gamma-k-def}.

The extension of the previous formalism to dynamical systems is straightforward. 
Within It$\bar{\rm o}$'s discretization scheme, the regulator in $Z_k[j,\tilde j]$ is a 
$2\times 2$ matrix since it is quadratic in the fields $(\phi,\tilde\phi)$. 
However, the admissible regulator terms are constrained by two requirements: 
(i) causality that will be discussed later, and (ii) the fact that the action must be proportional to  $\tilde\phi$ see Section \ref{msdrdj}. The latter condition implies in particular that terms such as
\begin{equation}
   \int_{x,x',t,t'} R_k^{\phi\phi}(x-x',t-t')\, \phi(x,t)\phi(x',t')
\end{equation}

are not allowed. In the following, we choose to implement a regulator only in the 
$\phi\tilde\phi$ sector, with a regulator function proportional to $\delta(t-t')$. 
This choice provides a convenient way to preserve causality.

The modified Legendre transform is defined as
\begin{equation}
\begin{split}
\Gamma_k[\varphi,\tilde\varphi] + \log Z_k[j,\tilde j]
&=\int_{\bf x}\left(j({\bf x}) \varphi({\bf x})
+\tilde j({\bf x})\tilde\varphi({\bf x})\right)\\
&\quad -\frac{1}{2}\int_{\bf q} R_k(q)\varphi({\bf q})\tilde\varphi(-{\bf q}),
\end{split}
\label{gammak}
\end{equation}
where ${\bf x}=(x,t)$, ${\bf q}=(q,\omega)$, $\varphi_i({\bf x})=\langle \phi_i({\bf x})\rangle $ with $i=1,2$ correspond respectively to $\phi({\bf x})$ and $\tilde\phi({\bf x})$.
The corresponding exact flow equation reads
\begin{equation}
\begin{split}
     \partial_k \Gamma_k[\varphi]   &=
\frac{1}{2}\, {\rm Tr}\int_{{\bf x},{\bf x'}} 
\partial_k \hat{R}_k({\bf x}-{\bf x'})\cdot
\hat{G}_k[{\bf x},{\bf x'};\varphi_i] \\
    &= \frac{1}{2}\tilde{\partial}_{k}{\rm Tr}\int_{\bf x}\left[ \log\Big(\hat\Gamma^{(2)}_{k}+\hat R_{k}\Big)\right]({\bf x},{\bf x})
\end{split}
\label{dkgam}
\end{equation}
where $\tilde\partial_k$ is the $k$-derivative acting only on the $k$-dependence of $R_k$ (and not on $\Gamma^{(2)}_k$) and 
$\hat{G}_k \equiv \left[\hat{\Gamma}_k^{(2)}+\hat{R}_k\right]^{-1}$ is
the full field-dependent propagator. Here, $\hat R_k$ is the $2\times2$ matrix with off diagonal elements $R_k(x-x')\delta(t-t')$ and 
$\hat{\Gamma}_k^{(2)}$ denotes the $2\times2$ matrix whose elements are 
$\Gamma^{(2)}_{k;ij}$, defined as
\begin{equation}
\Gamma^{(n)}_{k;i_1,\dots,i_n} [\{{\bf x}_i \};\varphi_i]=
\frac{\delta^n \Gamma_k [\varphi]}
{\delta \varphi_{i_1}({\bf x}_1)\dots\delta \varphi_{i_n}({\bf x}_n)}.
\end{equation}

The correspondence between the vertices 
$\Gamma^{(n)}_{k;i_1,\dots,i_n} [\{{\bf x}_i \};\varphi_i]$ and
\begin{equation}
\Gamma^{(n,\tilde{n})}_{k} [\{{\bf x}_i \},\{{\bf x'}_j \};\varphi_i]=
\frac{\delta^{n+\tilde{n}}\Gamma _k [\varphi]}
{\delta \varphi({\bf x}_1)\dots\delta\tilde{\varphi}({\bf x'}_{\tilde{n}})}
\end{equation}
is given in Appendix~8.2 of \cite{general-framework}. 
The implementation of It$\bar{\rm o}$'s prescription within this formalism is also detailed 
in the same reference.

The most commonly used approximation scheme to solve the RG flow 
Eq.~\eqref{dkgam} is the derivative expansion (DE). This approximation consists in 
expanding the (modified) effective action $\Gamma_k$ defined in 
Eqs.~\eqref{gamma-k-def} and \eqref{gammak} as a power series in the (time and space) derivatives of the fields, either $\varphi$ or $\varphi$ and $\tilde\varphi$, and truncating this expansion at a given finite order.

At equilibrium where there is no time derivative, the DE consists in an expansion in the gradient of the field that reads at order two:
\begin{equation}   \Gamma_k^{\rm\, eq}[\varphi]=\int_x\left[U_k(\varphi(x) +\frac{1}{2}Z_k(\varphi(x))\left( \nabla\varphi\right)^2 +O\left(\nabla^4\right) \right].
    \label{ansatz-eq}
\end{equation}
Out of equilibrium, since time-reversal symmetry, Eq.~\eqref{trs}, mixes $\tilde\varphi$ and 
$\partial_t\varphi$, any {\it ansatz} for $\Gamma_k$ truncated at order $n$ in time  derivatives must also involve an expansion in powers of $\tilde\varphi$ up to the  same order plus one (because $\Gamma_k$ is proportional to $\tilde\varphi$). Moreover, since the dynamical exponent $z$ is close to 2 in the  present model, it is natural to perform a joint expansion in time and space  derivatives, allowing twice as many derivatives in space as in time.

For instance, at second order in the DE, the {\it ansatz} for $\Gamma_k$ can be written
\begin{equation}
\begin{split}
\Gamma_k[\varphi,\tilde\varphi]&=\int_{\bf x} 
\, X_k(\varphi)\left(\tilde\varphi\,\partial_t \varphi -\tilde\varphi^2\right)\\
&+\tilde\varphi\left(U_k'(\varphi)- 
Z_k(\varphi)\,\nabla^2\,\varphi 
-\frac{1}{2}\,\partial_\varphi Z_k(\varphi) (\nabla \varphi)^2 \right).
\end{split}
\label{anz}
\end{equation}
The initial conditions of the flow are 
$U_\Lambda'=V'$, $Z_\Lambda=1$, and $X_\Lambda=1$. 
The parameterization of the last term, linear in $\tilde\varphi$, in 
Eq.~\eqref{anz} is chosen to facilitate the comparison between the flows of the 
equilibrium effective potential and field renormalization defined in Eq.~\eqref{ansatz-eq}, and those of 
$U_k$ and $Z_k$ defined above (see \cite{general-framework}).

In the supersymmetric formalism, $\Gamma_k$ is a functional of the superfield $\Psi=\langle \Phi\rangle$ which involves both the mean of $\phi$ and $\tilde\phi$ and of the Grassmann fields $\eta$ and $\bar\eta$ defined in Eq.~\eqref{superfield}. The exact RG flow is similar to the equilibrium one up to the fact that the superfield has coordinates along the $\theta$ and $\bar\theta$ directions which encompasses the out-of-equilibrium aspect of the model, see \cite{general-framework} for details.

At order two in derivatives, the {\it ansatz} for $\Gamma_k$ writes:
\begin{equation}
\Gamma_k[\Psi] = \int_\varpi  \left\{ X_k(\Psi)\bar D\Psi D \Psi + \frac{1}{2}\, Z_k(\Psi)\, (\nabla \Psi)^2 + U_k(\Psi)\right\}.
\label{anza}
\end{equation}
where we have again chosen the same name for the effective potential $U_k$ and field renormalization $Z_k$ as in  Eqs.~\eqref{ansatz-eq} and \eqref{anz} and the same name for $X_k$ as in Eq.~\eqref{anz}.

It has been shown in \cite{general-framework, Canet_2007} that the flows of  $U_k$, $Z_k$ and $X_k$ defined in both Eqs.~\eqref{anz} and \eqref{anza} are identical and for $U_k$ and $Z_k$ are also identical to the flows of the equilibrium effective potentials and field renormalization defined in Eq.~\eqref{ansatz-eq}.

Let us now address the question of the relationship between supersymmetry and TRS.

\section{Supersymmetry versus time-reversal symmetry}\label{reality}

In the It$\bar{\rm o}$ formalism, time-reversal symmetry is implemented by the transformations given in Eq.~\eqref{trs}. We show below that supersymmetry alone is not sufficient to enforce TRI. An additional ``reality'' condition, formulated in superspace, must be imposed in order to ensure TRI.

To this end, it is convenient to first perform a change of coordinates in the $(t,\theta,\bar\theta)$ superspace . We recall that this superspace should not be viewed as a direct sum of the real time direction with the Grassmann directions $(\theta,\bar\theta)$ \cite{Aron_2010, general-framework}. Indeed, in superspace, the time coordinate may involve any even combination of $\theta$ and $\bar\theta$ with $t$, such as non-real terms of the form $\alpha t+\beta\,\theta\bar\theta$ with $\alpha$ and $\beta\in\mathbb{R}$.

We thus introduce the following change of coordinates in superspace:
\begin{equation}
  \varpi = (t,\theta,\bar\theta)\ \to\ \varpi' = (t'=t+\tfrac{1}{2}\theta\bar\theta,\theta,\bar\theta).
   \label{new-coordinates-superspace}
\end{equation}
In terms of these new coordinates, the covariant derivatives defined in Eq.~\eqref{cov-der} take the form
\begin{equation}
    D = \partial_\theta - \frac{1}{2}\bar\theta \partial_{t'},
\qquad
\bar D = \partial_{\bar\theta}- \frac{1}{2}\theta \partial_{t'},
\label{cov-der-new}
\end{equation}
which preserves the anti-commutator algebra of the covariant derivatives since:
\begin{equation}
    \{D,\bar D\} = -\partial_{ t'}.
\end{equation}

By definition, a superfield transforms as a scalar under this change of coordinates, i.e.,
$\Phi'(\varpi')=\Phi(\varpi)$. Explicitly,
\begin{align}
\Phi'(\varpi')&=
\phi'(t',x)
+ \eta'(t',x)\bar\theta
+ \theta\,\bar\eta'(t',x)
+ \theta\bar\theta\,\tilde{\phi}'(t',x)
\nonumber\\
\Phi(\varpi)&=
\phi(t',x)
+ \eta(t',x)\bar\theta
+ \theta\,\bar\eta(t',x)
\nonumber\\
&\quad
+ \theta\bar\theta\left(\tilde{\phi}(t',x) -\frac{1}{2}\partial_{t'}\phi(t',x)\right),
\label{superfield-new-frame}
\end{align}
from which we deduce that 
\begin{equation}
   \phi=\phi',\ \eta=\eta',\  \bar\eta=\bar\eta',\  \tilde\phi'=\tilde\phi-\tfrac{1}{2}\dot\phi, 
\label{fields-prime}
\end{equation}
 where the dot denotes a time derivative.

In these coordinates, it is natural to define a conjugation operation by
\begin{equation}
    \theta^*= \bar\theta,\qquad \bar\theta^*= -\theta,\qquad t'^*=-t',
    \label{star}
\end{equation}
such that
\begin{equation}
    D^*=\bar D,\qquad \bar D^*=-D,\qquad \partial_{t'}^*=-\partial_{t'},
\end{equation}
which is consistent with the algebra~\eqref{cov-der-new}.

Equation~\eqref{star} shows that the time-reversal transformation $T(t)=-t$ is implemented on $t'$ by the star conjugation, since $T(t')=-t'=t'^*$. This is not the case for $t$, as one finds $t^*=-t-\theta\bar\theta$, a direct consequence of the definitions of the star conjugation and of $t'$.

In the following, we restrict ourselves to ``real'' superfields $\Phi$, defined by the condition
\begin{equation}
    \left(\Phi(\varpi)\right)^*=\Phi(\varpi^*).
\end{equation}
This implies
\begin{equation}
    \left(\phi'(t',x)\right)^*=\phi'(-t',x),\qquad
    \left(\tilde\phi'(t',x)\right)^*=\tilde\phi'(-t',x).
    \label{field-conjugation}
\end{equation}
The component fields $\phi'$ and $\tilde\phi'$ are therefore real with respect to the star conjugation. The Grassmann fields are conjugate to each other under this operation, in the same way as the coordinates $\theta$ and $\bar\theta$.

Let us now check that the time-reversal transformation of Eq.~\eqref{trs} is implemented on $\phi'$ and $\tilde\phi'$ through the star conjugation:
\begin{align}
    T(\phi'(t'))&=\phi'(-t')=\left(\phi'(t')\right)^*,\nonumber\\
    T(\tilde\phi'(t'))&=\tilde\phi'(-t')=\left(\tilde\phi'(t')\right)^*.
    \label{trs-new-field}
\end{align}
Using Eqs.~\eqref{trs-new-field}, \eqref{fields-prime} and \eqref{field-conjugation}, we find
\begin{equation}
   T(\phi'(t'))= T(\phi(t'))= \phi(-t') = \phi'(-t')=\left(\phi'(t')\right)^*
\end{equation}
which proves the first equation of \eqref{trs-new-field}. For $\tilde\phi'$, we obtain
\begin{equation}
\begin{split}
   T(\tilde\phi'(t'))&=T\left(\tilde\phi(t')-\frac{1}{2}\dot\phi(t')\right)  \\
      &=\tilde\phi(-t')-\dot\phi(-t')+\frac{1}{2}\dot\phi(-t')\\
      &=\tilde\phi'(-t'),\\
      &=\left(\tilde\phi'(t')\right)^*.
\end{split}
\end{equation}
which proves the second equality of Eq.~\eqref{trs-new-field}.

We emphasize that the TRS transformations, Eq.~\eqref{trs} is also implemented by the star conjugation since  $t^*=-t-\theta\bar\theta$ implies that $\left(\tilde\phi(t)\right)^*= \tilde\phi(-t) -\dot\phi(-t)$. The advantage of the $(t',\theta,\bar\theta)$ coordinates is of course to make trivial the star conjugation on the response field.

From the above discussion and the fact that $\int d\bar\theta d\theta $ is real under the star conjugation, we conclude that invariance of the action under TRS requires not only supersymmetry but also reality under the star conjugation. 

Let us finally note two things. 

First, the supersymmetric action in Eq.~\eqref{actionsusy} involves the term $D\Phi \bar D\Phi$ which is the only nonvanishing term invariant under supersymmetry of lowest degree in the time derivative. Once written in terms of $\phi$ and $\tilde\phi$ it reads $-\tilde\phi^2+\tilde\phi\dot\phi$ while in terms of the $\phi'$ and $\tilde\phi'$ fields, it is $\tilde\phi'{}^2 -\frac{1}{4}\dot\phi'{}^2$. Both $\tilde\phi'{}^2$ and $\dot\phi'{}^2$ are time reversal symmetric once they are integrated over $t$, see Eq.~\eqref{trs-new-field}, and therefore TRS alone is not sufficient to select the supersymmetric linear combination of these two terms. The same problem is somewhat hidden in the other coordinate system if we admit that in It$\bar{\rm o}$ the action is proportional to $\tilde\phi$. However, if we relax this condition, we can  formulate the problem this way: starting from an action that involves  $\tilde\phi^2$, $\tilde\phi \dot\phi$ and $\dot\phi^2$, can we find the supersymmetric extension of this action and does it select the term $-\tilde\phi^2+\tilde\phi\dot\phi$ for the part of the action independent of the Grassmann fields?

Using the supersymmetry generators defined in Eq.~\eqref{generators}, we find:
\begin{equation}
    \begin{split}
&Q \phi= \bar\eta\ ,\ \  \bar Q\phi=-\eta\\
& Q \eta= \phi\ ,\ \      \bar Q\eta=0\\
&Q\tilde\phi=0\ ,\ \      \bar Q\tilde\phi=-\dot\eta\\
 &Q\bar\eta=0\ ,\ \  \bar Q\bar\eta=\tilde\phi - \dot\phi.
    \end{split}
    \label{susy-transfo}
\end{equation}
On dimensional grounds, the action quadratic in the fields can only be:
\begin{equation}
   S=\int_{t,x}\left( \alpha\tilde\phi^2+ \beta\,\tilde\phi\dot\phi+\gamma\, \dot\phi^2+ \delta\,\dot\eta\bar\eta\right) 
\end{equation}
with $\alpha,\beta,\gamma$ and $\delta$ four real numbers. Requiring that the above action is invariant under the supersymmetry transformations Eq.~\eqref{susy-transfo}, we find that $\gamma=0$ and $\alpha=-\beta=-\delta$. The resulting action is therefore identical to the action in Eq.~\eqref{action-Ito} except for the Grassmann fields that, being decoupled from $\phi$ and $\tilde\phi$ play no physical role. We conclude that in terms of $\phi$ and $\tilde\phi$, supersymmetry is necessary to discard in the action the term not proportional to $\tilde\phi$ and to find the usual quadratic part of the TRS invariant action, Eq.~\eqref{action-Ito}.

Second, we can notice that in the above argument, the reality of the action has never been invoked somewhat at odd with our previous claim that it is a necessary ingredient to get TRS. The reason for this paradox is simple: at this order in $D$ and $\bar D$ the only non-vanishing term -- which is $D\Phi\bar D\Phi$ -- is real: there is no supersymmetric non-real term at this order. However, this is no longer true at the next orders of the DE where there exist supersymmetric terms that are not real. In this case, it is important to discard these terms in the effective action if we want to describe a system at thermal equilibrium. For instance, the most general supersymmetric ansatz at the next order of the DE can be written:
\begin{align}
\label{SM2}
     \Gamma_{k}[\Psi]  = \int_{\varpi}  \Big(&X_\kappa(\Psi)\bar D\Psi D \Psi + \frac{1}{2}\, Z_\kappa(\Psi)\, (\nabla \Psi)^2 + U_\kappa(\Psi)\nonumber\\
     &+B_{k}(\Psi)\left(\left(\bar{D}D\Psi\right)^2+\left(D\bar{D}\Psi\right)^2\right)\nonumber\\
     &+C_{k}(\Psi)\Psi\left(\left(\bar{D}D\right)^2+\left(D\bar{D}\right)^2\right)\Psi\nonumber\\
     &+ D_{k}(\Psi)\left(\left(\bar{D}D\Psi\right)^2-\left(D\bar{D}\Psi\right)^2\right)\Big)
 \end{align}
and it is not time reversal symmetric since its last term, proportional to $D_{k}(\Psi)$,  is not real. Rewritten in the  It$\bar{\rm o}$ formalism, $\Gamma_k$ is given by:
\begin{align}
\label{itoAT}
    \Gamma_{k}[\varphi, \tilde{\varphi}]  = \int_{t,x}
    &X_{k}(\varphi)(\tilde{\varphi}\partial_{t}\varphi-\tilde{\varphi}^2)+\tilde\varphi\Big(U_k'(\varphi)\nonumber\\
 &   - Z_k(\varphi)\,\nabla^2\,\varphi 
-\frac{1}{2}\, Z'_k(\varphi) (\nabla \varphi)^2 \Big)\nonumber\\
&- B'_{k}(\varphi)\left(\tilde{\varphi}(\partial_{t}\varphi)^2-3\tilde{\varphi}^2(\partial_{t}\varphi)+2\tilde{\varphi}^3\right)\nonumber\\
    & +\partial_{\varphi}\left(C_{k}(\varphi)\varphi\right)\left(\tilde{\varphi}\partial ^2_{t}\varphi-\partial_{t}\tilde{\varphi}\partial_{t}{\varphi}\right)\nonumber\\
    &- D'_{k}(\varphi)\left(\tilde{\varphi}\left(\partial_{t}\varphi\right)^2-\tilde{\varphi}^2\left(\partial_{t}\varphi\right)\right),
\end{align}
and the last term in Eq.~\eqref{itoAT} is not TRI  since it is  anti-symmetric under the time reversal transformations of Eq.~\eqref{trs}. 

Let us now demonstrate that one can construct a Langevin equation whose associated MSRDJ action is supersymmetric while explicitly breaking TRI.

For simplicity, we adopt a constructive approach starting from the action, from which the corresponding Langevin dynamics can be inferred. The supersymmetric non-real action we consider reads
\begin{equation}
    \begin{split}
    S &= \int_\varpi \Big( H(\Phi) + \bar D \Phi\, D \Phi 
    + a \Phi^2 \big[(D \bar D \Phi)^2 - (\bar D D \Phi)^2\big] \\
    &\qquad\qquad\quad + a^2 \Phi^2 (D \bar D \Phi + \bar D D \Phi)^2 \bar D \Phi\, D \Phi \Big) \\
    &= \int_\varpi \Big( H(\Phi) + \bar D \Phi\, D \Phi 
    \big[1 + a \Phi (D \bar D \Phi + \bar D D \Phi)\big]^2 \Big).
    \end{split}
\end{equation}

Expressed in terms of the fields $\phi$ and $\tilde{\phi}$, and after integrating out the decoupled Grassmann variables, this action reduces to
\begin{equation}
    S = \int_{\mathbf{x}} \tilde{\phi} \, (\dot{\phi} - \tilde{\phi}) (1 - a \phi \dot{\phi})^2 
    + \tilde{\phi} \left( -\nabla^2 \phi + V'(\phi) \right).
\end{equation}

The deterministic contribution to the Langevin equation associated with this action is obtained from the term linear in $\tilde{\phi}$, yielding
\begin{equation}
    -\nabla^2 \phi + V'(\phi) + \dot{\phi} (1 - a \phi \dot{\phi})^2.
\end{equation}
The noise term, on the other hand, follows from the term quadratic in $\tilde{\phi}$. Altogether, the Langevin equation associated with the above MSRDJ action takes the form
\begin{equation}
    \dot{\phi} (1 - a \phi \dot{\phi})^2 
    = \nabla^2 \phi - V'(\phi) + (1 - a \phi \dot{\phi}) \, \zeta,
    \label{langevin-non-trs}
\end{equation}
where $\zeta$ is a Gaussian white noise of zero mean and variance defined in Eqs.~\eqref{noise}.

In summary, we have constructed a field theory that is supersymmetric but not invariant under time reversal. This theory captures the long-time statistical properties of a system governed by the Langevin dynamics \eqref{langevin-non-trs}, provided that the system reaches a stationary state independent of its initial condition.

Although the stationary dynamics explicitly breaks TRI, one may conjecture that an effective restoration of this symmetry emerges at large scales. Indeed, the terms proportional to $a$ in Eq.~\eqref{langevin-non-trs} are expected to be strongly irrelevant in the renormalization group sense. In the absence of a nonperturbative fixed point, this suggests that, for instance near a second-order phase transition, the model defined by Eq.~\eqref{langevin-non-trs} should belong to the same universality class as the equilibrium Ising model for static properties, and share the same dynamical critical exponent $z$ as Model A.

An intriguing alternative scenario would be the existence of a nonperturbative fixed point -- either in this model or in a broader class of supersymmetric but time-reversal-breaking theories -- at which the time-reversal symmetry breaking terms do not vanish. This would define a new universality class. Such a possibility could in principle be investigated using the functional renormalization group.

\section{Decoupling between equilibrium and dynamical flows}\label{decoup}

For systems obeying detailed balance, the long-time limit corresponds to thermal equilibrium. In this regime, the probability distribution of microstates -- the Gibbs distribution -- is independent of the underlying dynamics. As a consequence, all quantities defined at equilibrium are dynamics-independent, and in particular so are the equilibrium critical exponents.

More generally, one expects that the renormalization-group flows of the equilibrium functions -- namely those entering the derivative expansion of $\Gamma_k^{\rm\, eq}[\varphi]$, see Eq.~\eqref{ansatz-eq} -- are decoupled from the other functions that we call the  dynamical functions. The latter correspond to terms in $\Gamma_k[\varphi,\tilde{\varphi}]$ involving time derivatives $\partial_t^n \varphi$ with $n \ge 1$, or to terms proportional to $\tilde{\varphi}^n$ with $n>1$, such as the functions $X_k(\varphi)$, $B_k(\varphi)$, and $C_k(\varphi)$ appearing in Eq.~\eqref{itoAT}.

Our aim is to make explicit how this decoupling emerges within the derivative expansion of the RG flow of $\Gamma_k[\varphi,\tilde{\varphi}]$. We show below that TRI is not necessary for this decoupling, only supersymmetry is. This leaves the intriguing possibility of having systems that explicitly break TRI while showing the same static properties as time-reversal invariant systems.

For the sake of simplicity, we present the proof below up to second order in time derivatives of $\varphi$. This order already captures the essential mechanism underlying the decoupling. The complete proof, valid to all orders in the derivative expansion, is provided in Appendix \ref{appenA}.
Note that the proof at order $\partial_t \varphi$ is too elementary to reveal the generic decoupling mechanism, which motivates working at second order.

We start by considering theories showing TRI. Within the It\={o} formalism, we first consider the most general TRI form of $\Gamma_{k}[\varphi, \tilde{\varphi}]$, expanded up to second order in time derivatives:
\begin{equation}
    \begin{split}
 \Gamma_{k}[\varphi, \tilde{\varphi}] =& \int_{\mathbf{x}}\Big( X_{k}(\varphi)\big(\tilde{\varphi}\,\partial_{t}\varphi - \tilde{\varphi}^2\big) + \tilde{\varphi}\,F_{k}[\varphi;\mathbf{x}] \\
 &+ B_{k}(\varphi)\big(\tilde{\varphi}(\partial_{t}\varphi)^2 - 3\tilde{\varphi}^2(\partial_{t}\varphi) + 2\tilde{\varphi}^3\big) \\
 &+ C_{k}(\varphi)\big(\tilde{\varphi}\,\partial_{t}^2\varphi - (\partial_{t}\tilde{\varphi})(\partial_{t}\varphi)\big) \Big),
    \end{split}
    \label{ansatz-order-2}
\end{equation}
where $F_{k}[\varphi;\mathbf{x}]=F_k\big(\varphi(x,t),\nabla\varphi(x,t),\nabla^2\varphi(x,t),\ldots\big)$ depends on $\varphi$ and on all its spatial derivatives (see Eq.~\eqref{itoAT} for the first terms in its gradient expansion). In Eq.~\eqref{ansatz-order-2}, the functions $B_k$, $C_k$, and $X_k$ are arbitrary functions of $\varphi(x,t)$ and correspond to redefinitions of those appearing in Eq.~\eqref{itoAT}. As mentioned above, the expansion in $\partial_t$ also amounts to an expansion in powers of $\tilde{\varphi}$, since time-reversal transformations mix $\partial_t \varphi$ and $\tilde{\varphi}$.

Our goal is to show that the RG flow of $F_k$ is closed, i.e., that it depends only on $F_k$ itself and not on the other functions entering $\Gamma_k[\varphi,\tilde{\varphi}]$. The functional $F_k$ is defined by 
\begin{equation}
    F_{k}[\varphi;x] = \left.\frac{\delta \Gamma_k[\varphi,\tilde{\varphi}]}{\delta \tilde{\varphi}(x,t)}\right|_{\substack{\varphi=\varphi(x) \\ \tilde{\varphi}=0}},
\label{def-F}
\end{equation}
where the right-hand side is evaluated for a time-independent field configuration $\varphi(x)$. This eliminates all terms proportional to $\partial_t \varphi$ and projects onto the contribution proportional to $F_k$.

Using Eq.~\eqref{dkgam}, the flow of $F_k$, written in position and time space, reads 
\begin{equation}
\begin{split}
  \partial_{k}F_{k}[\varphi;x]= &\frac{1}{2}\,\tilde{\partial}_{k}
 \mathrm{Tr}\int_{\mathbf{u},\mathbf{v}}\Bigg(
\left(\hat\Gamma^{(2)}_{k}+\hat R_{k}\right)^{-1}[\mathbf{u},\mathbf{v};\varphi,\tilde\varphi]
 \\
&\qquad\left.\cdot\frac{\delta\hat\Gamma^{(2)}_{k}[\mathbf{v},\mathbf{u};\varphi,\tilde\varphi]}{\delta\tilde{\varphi}(x,t)}\Bigg)
\right|_{\substack{\varphi=\varphi(x)\\ \tilde{\varphi}=0}}.  
\end{split}
\label{flow-F}
\end{equation}
It can equivalently be written as
\begin{equation}
  \partial_{k}F_{k}[\varphi;x]= -\frac{1}{2}
 \mathrm{Tr}\,\Bigg(
\hat G_k\cdot\partial_k\hat R_k\cdot\hat G_k
 \left.\cdot\frac{\delta\hat\Gamma^{(2)}_{k}}{\delta\tilde{\varphi}(x,t)}\Bigg)
\right|_{\substack{\varphi=\varphi(x)\\ \tilde{\varphi}=0}},  
\label{flow-F-bis}
\end{equation}
where $\hat G_k=\left(\hat\Gamma^{(2)}_{k}+\hat R_{k}\right)^{-1}$ is the full propagator. Here, $\mathrm{Tr}$ denotes both the trace over the $2\times2$ matrix structure and the integration over all intermediate positions, with the first and last points identified:
\begin{equation}
\mathrm{Tr}\,A_1\cdots A_n=\int_{x_1,\ldots,x_n}A_1(x_1,x_2)\cdots A_n(x_n,x_1).
\end{equation}
Although the second form, Eq.~\eqref{flow-F-bis}, is algebraically more involved, it leads to frequency integrals that are better behaved at large $\omega$. In the following, we make use of both representations, Eqs.~\eqref{flow-F} and \eqref{flow-F-bis}.

The flow of $F_k$ is computed in a standard way. We first determine the propagator $(\hat\Gamma^{(2)}_{k}+\hat R_{k})^{-1}$ and the vertex function $\delta\hat\Gamma^{(2)}_{k}/\delta\tilde{\varphi}$ from the ansatz~\eqref{ansatz-order-2}. We then evaluate these quantities at $\tilde\varphi=0$ and for a time-independent field configuration $\varphi(x)$, and finally perform the frequency integral arising in the trace.

We start by computing the propagator in the limit $\tilde{\varphi}=0$. A key point is that the term proportional to $C_{k}(\varphi)$ modifies the propagator. This is precisely why the calculation becomes nontrivial at second order in $\partial_t$ within the derivative expansion. In the following, we perform a Fourier transform in time. This is indeed convenient because the right-hand side of Eqs.~\eqref{flow-F} and \eqref{flow-F-bis} is evaluated for a time-independent configuration $\varphi(x)$ and is thus invariant under time translations. We obtain
\begin{equation}
\left.(\hat\Gamma^{(2)}_{k}+\hat R_{k})\right|_{\substack{\varphi(x)\\ \tilde{\varphi}=0}}=
 \begin{pmatrix}
0 & h_k(\omega)\\
h_k(-\omega) & \dfrac{1}{i\omega}\big(h_k(\omega)-h_k(-\omega)\big)
\end{pmatrix},
\label{gamma2+R-1}
\end{equation}
where $h_k(\omega)$ is a shorthand notation for
\begin{equation}
\label{hk}
\begin{split}
  h_k(\omega,&x-y,\varphi(y),\cdots, \nabla^n\varphi(y),\cdots)=\frac{\delta F_{k}[\varphi;\mathbf{y}]}{\delta \varphi(x)}+R_{k}(x-y) \\
  &+\Big(i\omega\, X_{k}(\varphi(y)) - 2\omega^2\, C_{k}(\varphi(y))\Big)\,\delta(x-y).
\end{split}
\end{equation}
The left-hand side of Eq.~\eqref{gamma2+R-1} is itself a function of $\omega$, $u-v$, and $\varphi(v)$. 

An important remark is in order here. In standard calculations within the It\={o} formalism, the propagator $\langle \tilde\varphi \varphi \rangle$ is usually endowed with a factor $e^{\pm i\epsilon \omega}$ in order to encode causality~\cite{general-framework}. This factor also regularizes the frequency integrals, and for any well-defined observable the limit $\epsilon \to 0$ can be safely taken at the end of the calculation. In the present case, however, such a regularization is not required. Indeed, the contribution proportional to $C_k$ is sufficient to render the frequency integrals convergent. As a consequence, the limit $\epsilon \to 0$ can be taken prior to performing the integration, since for convergent integrals, integrating over $\omega$ and taking the $\epsilon \to 0$ limit commutes.

The propagator, defined as the inverse of $\hat\Gamma^{(2)}_{k}+\hat R_{k}$, reads
\begin{equation}
(\hat\Gamma^{(2)}_{k}+\hat R_{k})^{-1}=
 \begin{pmatrix}
\frac{1}{i\omega}\big(h_k^{-1}(\omega)-h_k^{-1}(-\omega)\big) & h_k^{-1}(-\omega)\\
h_k^{-1}(\omega) & 0
\end{pmatrix},
\label{propag}
\end{equation}
where $h_k^{-1}(\omega)=h_k^{-1}(\omega,v-u,\varphi(v))$ denotes the inverse of $h_k(\omega)$ in the operator sense, that is, in position space:
\begin{equation}
    \int_y h_k(\omega,x-y,\varphi(x),\cdots)\, h_k^{-1}(\omega,y-z,\varphi(y),\cdots) = \delta(x-z).
\end{equation}

We now compute $\delta\hat\Gamma^{(2)}_{k}[\mathbf{u},\mathbf{v};\varphi,\tilde\varphi]/\delta\tilde{\varphi}(x,t)$ in the configuration where $\tilde{\varphi}=0$ and $\varphi=\varphi(x)$ is time-independent. In frequency space and at vanishing external frequency, we obtain:

\begin{widetext}
\begin{equation}
\label{gamma3}
\begin{split}
 &\left.\frac{\delta\hat\Gamma^{(2)}_{k}[\mathbf{u},\mathbf{v};\varphi,\tilde\varphi]}{\delta\tilde{\varphi}(x,t)}\right|_{\begin{array}{l}
\scriptstyle \varphi=\varphi(x)\\
\scriptstyle \tilde{\varphi}=0
\end{array}}\underset{\rm FT}{=}\\
&\begin{pmatrix}
\frac{\delta^2F_k[\varphi;x]}{\delta\varphi(u)\delta\varphi(v)}+2\left(B_k(\varphi(x))\delta(x-v)-\frac{\delta C_k(\varphi(x))}{\delta\varphi(v)}\right)\omega^2\delta(x-u)
&&\left(-2\frac{\delta X_k(\varphi(x))}{\delta \varphi(v)}+6B_{k}(\varphi(x))i\omega\delta(x-v)\right)\delta(x-u)\\
\left(-2\frac{\delta X_k(\varphi(x))}{\delta \varphi(v)}-6B_{k}(\varphi(x))i\omega\delta(x-v)\right)\delta(x-u)
&&12B_k(\varphi(x))\delta(x-u)\delta(x-v)
\end{pmatrix}.
\end{split}
\end{equation} 
\end{widetext}
Here, ``FT'' indicates that a Fourier transform with respect to time has been performed.

The trace of the product of the propagator and $\delta\hat\Gamma^{(2)}_{k}/\delta\tilde{\varphi}(x,t)$ can now be evaluated in Eq.~\eqref{propag}. The remaining crucial step is the integration over $\omega$. As we show below, it is precisely this integral, together with causality, that ensures that none of the dynamical functions -- here $X_k$, $B_k$, and $C_k$ -- contributes to the flow of $F_k$.

To make this decoupling explicit, we simplify the notation of $\delta\hat\Gamma^{(2)}_{k}/\delta\tilde{\varphi}(x,t)$ by suppressing its dependence on $x$, $u$, and $v$, which is irrelevant for the frequency integration. Equation~\eqref{gamma3} can then be written in the symbolic form
\begin{equation}
\frac{\delta\hat\Gamma^{(2)}_{k}}{\delta\tilde{\varphi}}\bigg|_{\tilde{\varphi}=0}=\begin{pmatrix}
F_{k}''+2\left(B_{k}-C'_{k}\right)\omega^2 & -2X'_{k}+6B_{k}i\omega\\
-2X'_{k}-6B_{k}i\omega & 12B_{k}
\end{pmatrix}.
\end{equation}

Inserting these expressions into Eq.~\eqref{flow-F}, we obtain
\begin{align}
\label{flow-F-ter}
&\partial_{k}F_{k}=\frac{1}{2}\tilde{\partial}_{k}\,\mathrm{Tr}\nonumber\\
&\int_{\omega} \Bigg(\frac{1}{i\omega}\left(h_k^{-1}(\omega) - h_k^{-1}(-\omega)\right)\cdot\big(F_k''+2(B_k-C'_k)\omega^2\big)\nonumber\\
&\qquad - 4 X'_k\cdot h_k^{-1}(\omega)+6 i\omega B_k\cdot\big(h_k^{-1}(\omega)-h_k^{-1}(-\omega)\big) 
\Bigg),
\end{align}
where the dot and the trace indicate that the product must be understood as an integral over space variables as in Eqs.~\eqref{flow-F} and \eqref{flow-F-bis}.

The integral over $\omega$ can now be performed by noting that $h_k^{-1}(-\omega)$ is the Fourier transform of the response function $\langle \tilde\phi(t') \phi(t)\rangle$, which is causal, i.e., proportional to $\theta(t-t')$. As a consequence, in the complex $\omega$-plane, all poles of this function lie in the lower half-plane. This property is preserved along the RG flow, provided that the regulator $R_k$ does not break causality~\cite{general-framework}, which is indeed the case for the frequency-independent regulator considered here. It follows that any convergent $\omega$-integral involving $h_k^{-1}(\pm \omega)$ multiplied by a polynomial in $\omega$ vanishes, since the integration contour can be closed by a semicircle at infinity without enclosing any pole. Note that if the regularizing factor $\exp(\pm i\epsilon\omega)$ is retained in the propagator, the contour must  be closed in one given half-plane according to the sign in this factor and this half-plane is precisely the one that does not contain poles.

A subtlety arises because the above argument applies only to convergent integrals. However, to make use of it, it is convenient to decompose the $\omega$-integral into separate contributions from each term in the integrand, which are not necessarily convergent individually. For instance, at second order in the derivative expansion, $h_k(\omega)\sim \omega^2$ (see Eq.~\eqref{hk}), so that $h_k^{-1}(\omega)\sim \omega^{-2}$. As a result, for the last term in Eq.~\eqref{flow-F-ter}, one has
\begin{equation}
\begin{split}
     &\tilde\partial_k\,{\rm Tr}\int_{\omega} \omega B_k\cdot\big(h_k^{-1}(\omega)-h_k^{-1}(-\omega)\big) \\
     &\neq  \tilde\partial_k\,{\rm Tr}\int_{\omega} \omega B_k\cdot h_k^{-1}(\omega)
     - \tilde\partial_k\,{\rm Tr}\int_{\omega} \omega B_k\cdot h_k^{-1}(-\omega).
\end{split}
\end{equation}

Two strategies can then be adopted. One may regularize the integral before splitting it into separate terms, which effectively amounts to reintroducing the factor $\exp(\pm i\epsilon\omega)$. Alternatively, one can first act with $\tilde\partial_k$, which corresponds to working with Eq.~\eqref{flow-F-bis} rather than Eq.~\eqref{flow-F}. The latter approach improves the convergence of the $\omega$-integrals, since
\[
\tilde\partial_k h_k^{-1}(\omega) = -\,h_k^{-1}(\omega)\cdot \partial_k R_k \cdot h_k^{-1}(\omega),
\]
which involves two propagators instead of one. In this case, both integrals
\[
{\rm Tr}\int_{\omega} \omega B_k\cdot h_k^{-1}(\pm\omega)\cdot \partial_k R_k\cdot h_k^{-1}(\pm\omega)
\]
are convergent, and the argument given above applies. The integration contour in the complex $\omega$-plane can then be closed by a semicircle at infinity without enclosing any pole, implying that both integrals vanish.

By inspection, one finds that the above argument applies to all terms in Eq.~\eqref{flow-F-ter} except the one proportional to $F_k''$. For this term, it is convenient to rewrite the integral as
\begin{equation}
\begin{split}
     &  \int_{\omega}\frac{1}{i\omega}\big(h_k^{-1}(\omega) - h_k^{-1}(-\omega)\big)\\
      & = \frac{1}{2}\lim_{\epsilon\to 0^+} \int_{\omega} \left(\frac{1}{i\omega+\epsilon}+\frac{1}{i\omega-\epsilon}\right)\big(h_k^{-1}(\omega) - h_k^{-1}(-\omega)\big),
\end{split}
\label{integral}
\end{equation}
which can now be decomposed into a sum of convergent integrals corresponding to the four terms obtained by expanding the integrand. Performing the change of variable $\omega\to -\omega$ in two of these terms, the integral in Eq.~\eqref{integral} can be rewritten as
\begin{equation}
   \lim_{\epsilon\to 0^+} \int_{\omega} \frac{h_k^{-1}(\omega)}{i\omega +\epsilon} + \lim_{\epsilon\to 0^+} \int_{\omega} \frac{h_k^{-1}(\omega)}{i\omega -\epsilon}.
\end{equation}
The second integral vanishes since all poles lie in the lower half-plane. For the first integral, the contour can be closed by a semicircle at infinity in the upper half-plane. Applying the residue theorem yields
\begin{equation}
\begin{split}
   \lim_{\epsilon\to 0^+} \int_{\omega} \frac{h_k^{-1}(\omega,u-v,\varphi(v),\cdots)}{i\omega +\epsilon}
   &= h_k^{-1}(\omega=0,u-v,\varphi(v),\cdots)\\
   &=\left(\frac{\delta F_{k}[\varphi;v]}{\delta \varphi(u)}+R_{k}(u-v)\right)^{-1}.
\end{split}  
\end{equation}

We thus arrive at
\begin{equation}
   \label{flow-F-final}
\partial_{k}F_{k}[\varphi,x]=\frac{1}{2}\tilde{\partial}_{k}\int_{u,v}\left(\frac{\delta F_{k}[\varphi;v]}{\delta \varphi(u)}+R_{k}(u-v)\right)^{-1}\cdot\frac{\delta^2F_k[\varphi;x]}{\delta\varphi(u)\delta\varphi(v)}.
\end{equation}
This equation is identical to the flow equation of $\delta\Gamma_k^{\rm \, eq}[\varphi]/\delta\varphi(x)$, where $\Gamma_k^{\rm \, eq}[\varphi]$ is the equilibrium functional defined in Eq.~\eqref{gamma-k-def}, whose flow is obtained by differentiating Eq.~\eqref{dkgam}. The identification $F_k[\varphi;x]=\delta\Gamma_k^{\rm \, eq}[\varphi]/\delta\varphi(x)$ is natural: the Langevin equation~\eqref{langevin} involves the functional derivative of the Hamiltonian $H$, which is the initial condition of the RG flow of $\Gamma_k^{\rm \, eq}$ at the ultraviolet scale $\Lambda$. Thus, $F_\Lambda[\varphi;x]=\delta H/\delta\varphi(x)$, and by construction this relation is preserved along the flow with $H$ replaced by $\Gamma_k^{\rm \, eq}$. This shows that $F_k[\varphi;x]=\delta\Gamma_k^{\rm \, eq}[\varphi]/\delta\varphi(x)$ at all scales.

To complete the proof at second order in the time-derivative expansion, we supplement the ansatz of Eq.~\eqref{ansatz-order-2} with the term that is antisymmetric under time-reversal transformation,
\begin{equation}
\label{Dk-term}
    D_k(\varphi)\left(\tilde{\varphi}\left(\partial_t \varphi\right)^2
-\tilde{\varphi}^2 \left(\partial_t \varphi\right)\right).
\end{equation}
The proof of decoupling presented above can then be repeated essentially unchanged. Indeed, this additional contribution does not modify the propagator, so that all previous arguments remain valid. This demonstrates that decoupling still holds at this order of the derivative expansion even in the absence of time-reversal invariance, provided that the model retains supersymmetry.

We have therefore shown, at second order in the time-derivative expansion of $\Gamma_k[\varphi,\tilde\varphi]$, that the RG flow of $F_k(\varphi(x),\nabla\varphi(x),\ldots)$ is independent of the  dynamical functions entering $\Gamma_k[\varphi,\tilde\varphi]$, and that it coincides with the flow of (the derivative of) the equilibrium functional $\Gamma_k^{\rm \, eq}[\varphi]$. In Appendix~\ref{appenA}, we extend this result to all orders in the time-derivative expansion of $\Gamma_k[\varphi,\tilde\varphi]$.

\section{Rate function for model A}

Not all universal critical quantities of the equilibrium Ising model can be computed from $\Gamma_k^{\rm \, eq}$. A well-known example is the probability distribution function (PDF) of the total spin of the system, $\sum_i \sigma_i/N$, where $N$ is the number of lattice sites and $\sigma_i=\pm1$ denotes the Ising spin at site $i$.

At criticality, this PDF is universal and can be computed in the continuum limit from the $\phi^4$ model. In a finite size system of linear size $L$, it reads
\begin{equation}
\label{pdf}
     P^{\rm\, eq}\left(L^{-d}\int_x\phi(x)=s\right)={\cal N}\int D\phi\, \delta\left(L^{-d}\int_x\phi(x)-s\right) \,e^{-H[\phi]}
\end{equation}
where ${\cal N}$ is a normalization factor and the $\beta=1/k_BT$ factor has been omitted to alleviate notation. When periodic boundary conditions are imposed, its calculation has been carried out in dimension three by  using either the functional RG~\cite{balog2022critical, 10.21468/SciPostPhys.18.4.119, PhysRevE.111.034131, FRA} or the perturbative RG within the $\epsilon$-expansion~\cite{sahu2024generalization, Sahu_2025, Sahu:2025fou, eisenriegler1987helmholtz, Rudnick1985, Rudnick1998}. 

This computation relies on the fact that the logarithm of the PDF, known as the rate function, can be expressed as the (modified) Gibbs free energy of a model for which the ``total spin'', $\int_x\phi(x)$, is constrained. More precisely, in the continuum, freezing this mode to a prescribed value can be implemented by modifying the Hamiltonian of the model in a way such that the fluctuations of $\int_x\phi(x)$ are suppressed. Defining
\begin{equation}
\label{ZM}
    Z_M^{\rm eq}[j]= \int D\phi\, e^{-H_M[\phi]+\int_x j\phi},
\end{equation}
with 
\begin{equation}
    H_M[\phi]=H[\phi]+ \frac{M^2}{2}\left(\int_x\big(\phi(x)-s\big)\right)^2,
    \label{HM}
\end{equation}
one can show that 
\begin{equation}
\begin{split}
P^{\rm\, eq}\left(L^{-d}\int_x\phi(x)=s\right)&\propto \lim_{M\to\infty}Z_M^{\rm eq}[j=0]\\
&\propto \lim_{M\to\infty} e^{-\Gamma_M^{\rm eq}[\varphi(x)=s]},
\end{split}
\label{P-versus-gamma-eq}
\end{equation}
where $\Gamma_M$ is defined through a (slightly modified) Legendre transform of $\log Z_M[j]$:
\begin{equation}
    \Gamma_M^{\rm\, eq}[\varphi]+\log Z_M[j]= j\cdot \varphi -\frac{M^2}{2}\left(\int_x\big(\varphi(x)-s\big)\right)^2.
\end{equation}
The rate function, defined as $I(s)=-L^{-d} \log P^{\rm\, eq}(s)$, is therefore given by
\begin{equation}
 I(s)=\lim_{M\to\infty}L^{-d}\Gamma_M^{\rm\, eq}[\varphi(x)=s]   
 \label{rate-function}
\end{equation}
in the  $L\to\infty$ limit.

An RG equation for $I(s)$ can be derived by following the same reasoning as in Eq.~\eqref{modified-H}, but now modifying $H_M$ instead of $H$:
\begin{equation}
    H_M\to H_{M,k}=H_M+\Delta H_k
    \label{HMk}
\end{equation}
 with $\Delta H_k$ defined in Eq.~\eqref{modified-H}. The partition function $Z_M$ is accordingly replaced by $Z_{M,k}$, and the running modified Gibbs free energy becomes
\begin{equation}
   \Gamma_{M,k}^{\rm\, eq}[\varphi] +\log Z_{M,k}[j]= j\cdot \varphi -\frac{1}{2}\varphi\cdot R_k\cdot \varphi -\frac{M^2}{2}\left(\int_x\big(\varphi(x)-s\big)\right)^2
\end{equation}
such that when $k=0$, $\Gamma_{M,k=0}^{\rm\, eq}[\varphi]=\Gamma_M^{\rm\, eq}[\varphi]$.

An exact RG flow equation for $\Gamma_{M,k}^{\rm\, eq}[\varphi]$ follows from this definition. It is identical to the usual flow equation~\eqref{dkgam}, up to the replacement $R_k(x-y)\to R_{M,k}(x-y)=R_k(x-y)+M^2$. 

Performing a derivative expansion of $ \Gamma_{M,k}^{\rm\, eq}[\varphi]$ as in Eq.~\eqref{ansatz-eq}, we find that the first term of this expansion, analogous to the derivative of the effective potential $U_k(\varphi)$, is the running  rate function $I_k(s)$ since up to a volume factor, it is by definition $ \Gamma_{M,k}^{\rm\, eq}[\varphi]$ evaluated in a constant field configuration $\varphi(x)=s$:
\begin{equation}
  I_k(s)=\lim_{M\to\infty}L^{-d}\Gamma_{M,k}^{\rm\, eq}[\varphi(x)=s] .
  \label{running-rate}
\end{equation}
 Its flow can be computed either perturbatively or within the derivative expansion, with the important subtlety that one must first work at finite volume and finite bulk correlation length $\xi_\infty$ (the correlation length of the infinite system), and then take the limits $L\to\infty$ and $\xi_\infty\to\infty$ at fixed ratio $\zeta=L/\xi_\infty$. In this limit, one finds a family of universal PDFs which depend on $\zeta$ and on the phase in which the system is when the critical limit $\xi_\infty\to\infty$ is taken \cite{balog2022critical, sahu2024generalization}: strictly speaking,  Eq.~\eqref{rate-function} should be replaced by: $I_\zeta(s)=\lim_{M\to\infty}L^{-d}\Gamma_M^{\rm\, eq}[\varphi(x)=s,\zeta]$ in the limit $L,\xi_\infty\to\infty$ at fixed ratio $\zeta$. Since we are not interested in the $\zeta$-dependence of the rate function in the following, we remove the index $\zeta$ in $I_\zeta(s)$ to alleviate the notation.

Our aim is to show that the rate function at equilibrium can be recovered from model A.   
 We show below that, by first taking the long-time limit in a modified model A, one can define a quantity that coincides with the rate function at equilibrium, and that this quantity follows the same RG flow as the equilibrium rate function. This provides an indirect proof that the equilibrium rate functions $P^{\rm\, eq}(s)$ can indeed be computed from model A.

The  calculation of the  rate function at equilibrium shows that computing the PDFs $P(s)$, see Eq.~\eqref{pdf}, is equivalent to computing the partition function -- more precisely, the modified Legendre transform of its logarithm -- of a model in which the total spin $\int_x \phi(x)$ is constrained to a fixed value. Motivated by this observation, we propose to modify the Langevin equation~\eqref{langevin} by adding a term analogous to the one proportional to $M^2$ in $H_M$, Eq.~\eqref{HM}:
\begin{equation}
    \partial_t \phi(x,t) = -\frac{\delta H_{M}[\phi]}{\delta \phi(x,t)} + \zeta(x,t).
    \label{langevin-M}
\end{equation}
The supplementary term in this equation, proportional to $M^2$, corresponds to an infinite force that pins the total spin to the value $s$.
Since this Langevin equation still belongs to the model A class, it follows that at finite size and in the long-time limit the system reaches thermal equilibrium, with a probability distribution of microstates given by the Gibbs measure $\exp\big(-H_{M}[\phi]\big)/Z_{M}$.

The same reasoning applies to the regularized Hamiltonian $H_{M,k}$ defined in Eq.~\eqref{HMk}. Its equilibrium dynamics can be described by the MSRDJ action associated with this model, obtained by adding to the action $S$ defined in Eq.~\eqref{action-Ito} the term
\begin{equation}
    \Delta S_{M,k}=\int_{\mathbf{x},\mathbf{y}}\tilde\phi(\mathbf{x})\Big[\big(R_k(x-y) +M^2 \big)\delta(t_x-t_y)\Big]\phi(\mathbf{y}).
    \label{deltaS-Mk}
\end{equation}
Proceeding as in the standard model A  with the action $S$, one defines a functional $\Gamma_{M,k}[\varphi,\tilde\varphi]$ associated with the modified action $S+\Delta S_{M,k}$, and all previous arguments can be repeated. In particular, its time-derivative expansion contains a term of the form $\tilde\varphi\, F_{M,k}[\varphi]$, analogous to the term $\tilde\varphi F_k[\varphi]$ in Eq.~\eqref{ansatz-order-2}. Evaluated for a uniform and time-independent field configuration $\varphi(x,t)=s$, the function $F_{M,k}(\varphi=s)$ corresponds to the dynamical counterpart of the (derivative of the) running rate function. 

The argument established above in Section \ref{decoup} for the flow of $F_k$ can then be extended to $F_{M,k}$, with the same conclusion: its RG flow is identical to the equilibrium one. In particular, this shows that the flow of the dynamical counterpart of the running (derivative of the) rate function coincides with that of the (derivative of the) equilibrium running rate function $I'_k(s)$ defined in Eq.~\eqref{running-rate}. Since for $k=0$ and $M\to\infty$ this running quantity reduces to the physical (derivative of the) rate function, we conclude that the rate function $I(s)$ -- and more generally all equilibrium rate functions $I_\zeta(s)$ -- can be computed within model A from  $F_{M,k=0}$  evaluated in the field configuration $\phi(x,t)=s$.

\section{Conclusion}
We have shown in this work that supersymmetry alone is not sufficient to guarantee time-reversal invariance, and we have provided an explicit example of a model exhibiting this property. We have further demonstrated that, provided such supersymmetric but non-time-reversal-invariant models relax toward a stationary state independent of the initial condition, all their static properties -- namely, all equal-time correlation functions evaluated in the stationary state -- coincide with those of a time-reversal-invariant model sharing the same deterministic dynamics. In particular, they possess the same critical exponents and, more generally, belong to the same static universality class.
Their dynamical properties, however, differ from those of time-reversal-invariant systems. Nevertheless, since the terms responsible for breaking time-reversal invariance are perturbatively irrelevant -- such as the $D_k$ term in Eq.~\eqref{Dk-term} -- and provided that the systems under consideration do not exhibit genuinely nonperturbative effects, such as nonperturbative fixed points of the renormalization-group flow \cite{PhysRevLett.95.100601, Canet_2025}, we expect these terms to become asymptotically irrelevant for correlation and response functions evaluated at sufficiently separated times.

We have also shown that the equilibrium probability distribution function $P(s)$ of the total spin,
\[
P\!\left(L^{-d}\!\int_x \phi(x)=s\right),
\]
can be computed directly within the Model A framework, without explicitly invoking the Gibbs equilibrium measure. This result relies on the fact that the corresponding PDF can be obtained from a modified Langevin dynamics in which the total spin is constrained to remain equal to $s$ throughout the time evolution. For systems that do not satisfy detailed balance, such a construction is no longer available, and our argument therefore does not extend to genuinely dynamical systems. Determining how to compute this quantity out of equilibrium thus remains an open problem.

Finally, let us emphasize that even within Model A, the computation of the critical distribution functions $P(s)$ remains highly nontrivial because the three limits -- infinite time, infinite volume, and approach to criticality -- do not commute. Already at equilibrium, the noncommutation of the last two limits implies the existence of not one but infinitely many universal critical PDFs, parametrized by the ratio $\zeta=L/\xi_\infty$, where $\xi_\infty$ is the bulk correlation length and both $L$ and $\xi_\infty$ are taken to infinity. We expect a similar situation to hold for Model A dynamics, where critical slowing down occurs at criticality. In this case, there should exist a doubly infinite family of universal critical PDFs, presumably parametrized by both $\zeta$ and $\zeta'=L^z/T$, where $T$ denotes the observation time and $z$ the dynamical critical exponent.

\appendix
\onecolumngrid
\section{Decoupling of the flow to all orders in derivative expansion in presence of TRI}\label{appenA}

In this section, we show how to generalize the results obtained in section \ref{decoup} and thereby show decoupling of the equilibrium part of the flow from its non-equilibrium counterpart up to all orders in the time derivative expansion. As shown in section \ref{reality}, along with supersymmetry an additional `reality' condition is an essential ingredient for imposing time-reversal symmetry in the Ito case. Using this, one can write down the form of the propagator in the supersymmetric language to all orders in the time derivative expansion which, at any order $p$ would now be given by:
\begin{equation}
    C^{(p)}_{k}(\Psi)\Psi\left((D\bar{D})^p+(-1)^p(\bar{D}D)^p\right)\Psi.
\end{equation}
Where $\Psi$ is a superfield defined in Eq.~\eqref{superfield}. This can be argued rather easily. For example, consider a term of the form $C^{(m+n)}_{k}(\Psi)\left((D\bar{D})^m\Psi(\bar{D}D)^n\Psi+(-1)^{(m+n)}(\bar{D}D)^m\Psi(D\bar{D})^n\Psi\right)$, which is the most generalized term one can write with two superfields respecting supersymmetry and the `reality' condition (with $m+n=p$). One can always perform integration by parts on this term to reach the term of the form $C^{p}_{k}(\Psi)\Psi\left((D\bar{D})^p+(-1)^p(\bar{D}D)^p\right)\Psi$ up to some additional terms containing three superfields which do not contribute to the propagator.\\

In the Ito-prescription (putting the Grassmann fields $\eta$ and $\bar{\eta}$ to $0$), thus one can write a generalized ansatz for $\Gamma_{k}$ given by:
\begin{align}\label{2DEN}
\Gamma_{k}[\varphi, \tilde{\varphi}] & = \int_{t,\vec{x}}\tilde{\varphi}F_{k}[\varphi; x]+\sum_{p\geq 1}\partial_{\varphi}\left(C^{(p)}_{k}(\varphi)\varphi\right)\left(\tilde{\varphi}\partial ^{p}_{t}\varphi-(-1)^p\partial_{t}{\varphi}\partial^{p-1}_{t}{\tilde{\varphi}}+((-1)^p-1)\tilde{\varphi}\partial^{p-1}_{t}\tilde{\varphi}\right)\nonumber\\
& +\sum_{m}J^{(m)}_{k}(\varphi)\tilde{\varphi}^{a}(\partial_{t}{\varphi})^{b}(\partial^{c}_{t}{\varphi}),
\end{align}
where the last term in Eq.~\eqref{2DEN} is a general term which must be included in accordance with time-reversal symmetry order by order in derivative-expansion (other than the propagator) with $m=a+b+c$. However, in the proof that follows we show that the term $\sum_{m}J^{(m)}_{k}(\varphi)\tilde{\varphi}^{a}(\partial_{t}{\varphi})^{b}(\partial^{c}_{t}{\varphi})$ is not important to prove the decoupling of the flow since this is linked to the structure of the propagator which can be exactly written down as seen in Eq.~\eqref{2DEN}. Using this ansatz, the inverse-propagator in Fourier space is thus given by (quite in the same spirit as its predecessor in section \ref{decoup}.):
\begin{equation}
\left.(\hat\Gamma^{(2)}_{k}+\hat R_{k})\right|_{\begin{array}{l}
\scriptstyle \varphi(x)\\
\scriptstyle \tilde{\varphi}=0
\end{array}}=
 \begin{pmatrix}
0&&\mathbf{h}(\omega)\\
\mathbf{h}(-\omega)&&\frac{1}{i\omega}(\mathbf{h}(\omega)-\mathbf{h}(-\omega))
\end{pmatrix}
\label{gamma2+R}
\end{equation} 
where $\mathbf{h}_k(\omega)$ is a simplified notation for 
\begin{equation}
\label{hk1}
\begin{split}
 \mathbf{h}_k(\omega,&x-y,\varphi(y),\cdots, \nabla^n\varphi(y),\cdots)=\frac{\delta F_{k}[\varphi; v]}{\delta \varphi(u)}+R_{k}(u-v)+\sum_{p\geq 1}2~(i\omega)^pC^{(p)}_{k}(\varphi(v))\delta(u-v) .
\end{split}
\end{equation}
Thus, the generalized propagator (up to any order in time derivative expansion) is given by :
\begin{equation}
(\hat\Gamma^{(2)}_{k}+\hat R_{k})^{-1}=
 \begin{pmatrix}
\frac{1}{i\omega}(\mathbf{h}_k^{-1}(\omega)-\mathbf{h}_k^{-1}(-\omega))&&\mathbf{h}_k^{-1}(-\omega)\\
\mathbf{h}_k^{-1}(\omega)&&0
\end{pmatrix}
\end{equation}

In the following, we consider the derivative expansion up to order $2n$ (in time, not time+space). For computing the flow of $F_{k}[\varphi; x]$, we also need to compute $\frac{\delta\Gamma^{(2)}_{k}}{\delta\tilde{\varphi}}$, which is given by:
\begin{align}
&\left.\frac{\delta\hat\Gamma^{(2)}_{k}[{ u},{ v};\varphi,\tilde\varphi]}{\delta\tilde{\varphi}(x,t)}\right|_{\begin{array}{l}
\scriptstyle \varphi=\varphi(x)\\
\scriptstyle \tilde{\varphi}=0
\end{array}}\underset{}{=}\begin{pmatrix}
\frac{\delta^2 F_{k}[\varphi; x]}{\delta\varphi(u)\delta\varphi(v)}+\sum_{p\geq 1}^{n} D^{(p)}_{k}(\varphi)\omega^{2p}&&\sum^{2n}_{p\geq 1}A^{(p)}_{k}(\varphi)(i\omega)^{p-1}\\
\sum^{2n}_{p\geq 1}A^{(p)}_{k}(\varphi)(-i\omega)^{p-1}&&J^{(3)}(\varphi)
\end{pmatrix}
\end{align}
Here, the coefficients $A^{(p)}_{k}(\varphi)$ and $D^{(p)}_{k}(\varphi)$ \footnote{The power of $\omega$ associated with $D^{(p)}_{k}(\varphi)$ will always be even,  since $\frac{\delta\Gamma^{(2)}_{k}}{\delta\tilde{\varphi}}$ is computed in a vanishing external frequency  configuration and any odd power in $\omega$ would be proportional to the external frequency.} appearing in front of $(i\omega)^{p}$ are obviously related to the coefficients $C^{(p)}_{k}(\varphi)$ and $J^{(m)}_{k}(\varphi)$ in some complicated fashion (that we don't really care about for the purposes of this proof). A subtlety which is however important to notice is that that the highest power of $\omega$ associated with $A^{(p)}_{k}(\varphi)$ is $\omega^{2n-1}$, since in the limit $\tilde{\varphi}=0$, the only non-vanishing contribution in $\frac{\delta\Gamma^{(2)}_{k}}{\delta\tilde{\varphi}}$ from terms upto order $2n$ in derivative-expansion would be given by terms of the structure: $\tilde{\varphi}^2\partial^{2n-1}_{t}(\varphi)$ (and other related terms). This is important since in an usual derivative expansion, one might expect that the highest power of $\omega$ associated with $A^{(p)}_{k}(\varphi)$ to be $\omega^{2n}$ but this doesn't come to fruition as this derivative expansion is a bit special since $\tilde{\varphi}$ plays the same role as $\partial_{t}$ in counting the order of this time-derivative expansion as explained in the core of the article. This also guarantees the convergence of the $\omega-$integrals order by order in the derivative expansion and alleviates the need to regularize the $\omega-$integral. \\

Using, the flow equation, we have:
\begin{align}\label{OmegaI}
\partial_{k}F_{k}[\varphi ; x] & = \frac{1}{2}\tilde{\partial_{k}}\int_{t,u, v}\Tr\left\{(\Gamma^{(2)}_{k}+R_{k})^{-1}.\frac{\delta\Gamma^{(2)}_{k}}{\delta\tilde{\varphi}}\right\}|_{\tilde{\varphi}=0, \varphi =\varphi(x)} \nonumber\\
& = \frac{1}{2}\tilde{\partial_{k}}\Tr\int_{\omega}\int_{u, v}\sum^{2n}_{p\geq 1}\mathbf{h}^{-1}_{k}(-\omega)\cdot A^{(p)}_{k}(\varphi)(-i\omega)^{p-1}+\mathbf{h}^{-1}_{k}(\omega)\cdot A^{(p)}_{k}(\varphi)(i\omega)^{p-1}\nonumber\\
& +\frac{1}{2}\tilde{\partial_{k}}\Tr\int_{\omega}\int_{u, v}\frac{1}{i\omega}\left(\mathbf{h}_k^{-1}(\omega)-\mathbf{h}_k^{-1}(-\omega)\right)\cdot\left(F''_{k}[\varphi; x]+\sum^{n}_{p\geq 1}D^{(p)}_{k}(\varphi)\omega^{2p}\right)
\end{align}
In Eq.~\eqref{OmegaI}, the $\omega-$integrals can be computed rather easily using residue theorem. We will analyze these integrals term by term. The first two-terms on the RHS of Eq.~\eqref{OmegaI}  can be computed in the following way:
\begin{align}
    & \frac{1}{2}\tilde{\partial_{k}}\Tr\int_{\omega}\int_{u, v}\sum^{2n}_{p\geq 1}\mathbf{h}^{-1}_{k}(-\omega)\cdot A^{(p)}_{k}(\varphi)(-i\omega)^{p-1}+\mathbf{h}^{-1}_{k}(\omega)\cdot A^{(p)}_{k}(\varphi)(i\omega)^{p-1}\nonumber\\
    & = \frac{1}{2}\tilde{\partial_{k}}\Tr\int_{\omega}\int_{u, v}\mathbf{h}^{-1}_{k}(-\omega)\cdot A^{(2n)}_{k}(\varphi)(-i\omega)^{2n-1}+\mathbf{h}^{-1}_{k}(\omega)\cdot A^{(2n)}_{k}(\varphi)(i\omega)^{2n-1}\nonumber\\
    & = -\frac{1}{2}\Tr\int_{\omega}\int_{u, v} \mathbf{h}_k^{-1}(-\omega)\cdot\partial_k R_k\cdot \mathbf{h}_k^{-1}(-\omega)\cdot A^{(2n)}_{k}(\varphi)(-i\omega)^{2n-1}+ \mathbf{h}_k^{-1}(\omega)\cdot\partial_k R_k\cdot \mathbf{h}_k^{-1}(\omega)\cdot A^{(2n)}_{k}(\varphi)(i\omega)^{2n-1}\nonumber\\
    & = 0
\end{align}
It must be noted that the integrals $\Tr\int_{\omega}\int_{u, v}\sum^{2n}_{p\geq 1}\mathbf{h}^{-1}_{k}(-\omega)\cdot A^{(p)}_{k}(\varphi)(-i\omega)^{p-1}$ and  $\Tr\int_{\omega}\int_{u, v}\sum^{2n}_{p\geq 1}\mathbf{h}^{-1}_{k}(\omega)\cdot A^{(p)}_{k}(\varphi)(i\omega)^{p-1}$ are not convergent by themselves but rather what is convergent is the combined sum of the two. More specifically, the individual convergence of each integral is explicitly broken by the highest power in $\omega$. This means that, if we consider the derivative expansion to order $2n$, the integrals $\Tr\int_{\omega}\int_{u, v}\sum^{2n-1}_{p\geq 1}\mathbf{h}^{-1}_{k}(-\omega)\cdot A^{(p)}_{k}(\varphi)(-i\omega)^{p-1}$ and  $\Tr\int_{\omega}\int_{u, v}\sum^{2n-1}_{p\geq 1}\mathbf{h}^{-1}_{k}(\omega)\cdot A^{(p)}_{k}(\varphi)(i\omega)^{p-1}$ will be individually convergent and their contributions will be vanishing, since by arguments of causality it can be shown that all the poles of $\mathbf{h}^{-1}_{k}(\pm\omega)$ will lie only in one-half of the complex $\omega-$plane which in turn, implies that all converging  $\omega$-integrals  involving $\mathbf{h}_k^{-1}(\pm\omega)$ multiplied by any polynomial in $\omega$ is vanishing since the contour of integration can be closed by a semi-circle at infinity such that no pole is enclosed in the contour. The integrals $\Tr\int_{\omega}\int_{u, v}\mathbf{h}^{-1}_{k}(-\omega)\cdot A^{(2n)}_{k}(\varphi)(-i\omega)^{2n-1}$ and $\Tr\int_{\omega}\int_{u, v}\sum^{2n}_{p\geq 1}\mathbf{h}^{-1}_{k}(\omega)\cdot A^{(p)}_{k}(\varphi)(i\omega)^{2n-1}$, which are not a priori individually convergent, can be made individually convergent by acting on it with the $\tilde{\partial}_{k}$. Once again, since the poles of $\mathbf{h}^{-1}_{k}(\pm \omega)$ lie only in one-half of the complex plane, these integrals vanish as well.
We now shift our focus to the rest of the terms in Eq.~\eqref{OmegaI}. Repeating similar arguments as above, one can directly conclude:
\begin{equation}
    \frac{1}{2}\tilde{\partial_{k}}\Tr\int_{\omega}\int_{u, v}\sum^{n}_{p\geq 1} (-1)^p\left(\mathbf{h}_k^{-1}(\omega)-\mathbf{h}_k^{-1}(-\omega)\right)\cdot D^{(p)}_{k}(\varphi)(i\omega)^{2p-1}=0
\end{equation}\\

Thus Eq.~\eqref{OmegaI} become :
\begin{equation}
    \partial_{k}F_{k}[\varphi]=\frac{1}{2}\tilde{\partial_{k}}\Tr\int_{\omega}\int_{u, v}\frac{1}{i\omega}\left(\mathbf{h}_k^{-1}(\omega)-\mathbf{h}_k^{-1}(-\omega)\right)\cdot F''_{k}[\varphi; x]
\end{equation}
Now, the last remaining $\omega$-integral, can be computed rather easily following the steps in section \ref{decoup} which gives:
\begin{align}\label{OmegaL}
    &\frac{1}{2}\tilde{\partial_{k}}\Tr\int_{\omega}\int_{u, v}\frac{1}{i\omega}\left(\mathbf{h}_k^{-1}(\omega)-\mathbf{h}_k^{-1}(-\omega)\right)\cdot F''_{k}[\varphi; x]\nonumber\\
    & = \frac{1}{2}\tilde{\partial_{k}}\Tr\int_{\omega}\int_{u, v}\mathbf{h}^{-1}_{k}(\omega=0)\cdot F''_{k}[\varphi; x]\nonumber\\
    & = \frac{1}{2}\tilde{\partial}_{k}\int_{\vec{u},\vec{v}}\left(\frac{\delta F_{k}[\varphi ; v]}{\delta \varphi(u)}+R_{k}(u-v)\right)^{-1}\cdot\frac{\delta^2 F_{k}[\varphi; x]}{\delta\varphi(u)\delta\varphi(v)},
\end{align}
where in the last line we have used the fact that:
\begin{equation}
    \mathbf{h}_k^{-1}(\omega=0,u-v,\varphi(v), \cdots)=\left(\frac{\delta F_{k}[\varphi ; v]}{\delta \varphi(u)}+R_{k}(u-v)+\sum_{p\geq 1}2~(i\omega)^pC^{(p)}_{k}(\varphi(v))\delta(u-v)\right)^{-1},
\end{equation}
from Eq.~\eqref{hk1}.
\section{Decoupling without time-reversal symmetry}
In this section, we show that we can have decoupling, even in absence of time-reversal symmetry. Essentially what we show here, is that the only requirement for the decoupling of the equilibrium flow from its non-equilibrium counterpart is guaranteed by the underlying supersymmetry. From now on, we relax the `reality'-condition that we discussed in the above section, i.e. In our FRG truncation, we also consider terms which are supersymmetric but unreal. This means that now, the generalized propagator will be of the form:
\begin{equation}
    C^{(p)}_{k}(\Psi)\Psi\left((D\bar{D})^p+(-1)^p(\bar{D}D)^p\right)\Psi+    M^{(p)}_{k}(\Psi)\Psi\left((D\bar{D})^p-(-1)^p(\bar{D}D)^p\right)\Psi,
\end{equation}
where $\Psi$ is a superfield defined in Eq.~\eqref{superfield}. Thus, in the Ito-prescription one can write a generalized ansatz for $\Gamma_{k}$ given by:
\begin{align}\label{2DENN}
\Gamma_{k}[\varphi, \tilde{\varphi}] & = \int_{t,\vec{x}}\tilde{\varphi}F_{k}[\varphi; x]+\sum_{p\geq 1}\partial_{\varphi}\left(C^{(p)}_{k}(\varphi)\varphi\right)\left(\tilde{\varphi}\partial ^{p}_{t}\varphi-(-1)^p\partial_{t}{\varphi}\partial^{p-1}_{t}{\tilde{\varphi}}+((-1)^p-1)\tilde{\varphi}\partial^{p-1}_{t}\tilde{\varphi}\right)\nonumber\\
& +\sum_{p\geq 1}\partial_{\varphi}\left(M^{(p)}_{k}(\varphi)\varphi\right)\left(\tilde{\varphi}\partial ^{p}_{t}\varphi+(-1)^p\partial_{t}{\varphi}\partial^{p-1}_{t}{\tilde{\varphi}}-((-1)^p+1)\tilde{\varphi}\partial^{p-1}_{t}\tilde{\varphi}\right)+\sum_{m}J^{(m)}_{k}(\varphi)\tilde{\varphi}^{a}(\partial_{t}{\varphi})^{b}(\partial^{c}_{t}{\varphi}),
\end{align}
where the last term in Eq.~\eqref{2DENN} is a general term which must be included in accordance with supersymmetry order by order in derivative-expansion (other than the propagator) with $m=a+b+c$.\\

The inverse-propagator in Fourier space will now be given by :
\begin{equation}
\left.(\hat\Gamma^{(2)}_{k}+\hat R_{k})\right|_{\begin{array}{l}
\scriptstyle \varphi(x)\\
\scriptstyle \tilde{\varphi}=0
\end{array}}=
 \begin{pmatrix}
0&&\mathbf{h}_k(\omega)\\
\mathbf{h}_k(-\omega)&&\frac{1}{i\omega}(\mathbf{h}_k(\omega)-\mathbf{h}_k(-\omega))+\sum_{p\geq 1}\partial_{\varphi}\left(M^{(p)}_{k}(\varphi)\varphi\right)(1+(-1)^p)(i\omega)^{p-1}
\end{pmatrix}
\label{gamma2+R'}
\end{equation} 
The corresponding propagator (up to any order in time derivative expansion) is now given by :
\begin{equation}
(\hat\Gamma^{(2)}_{k}+\hat R_{k})^{-1}=
 \begin{pmatrix}
\frac{1}{i\omega}(\mathbf{h}_k^{-1}(\omega)-\mathbf{h}_k^{-1}(-\omega))-\mathbf{h}_k^{-1}(-\omega)\cdot\left(\sum_{p\geq 1}\partial_{\varphi}\left(M^{(p)}_{k}(\varphi)\varphi\right)(1+(-1)^p)(i\omega)^{p-1}\right)\cdot \mathbf{h}_k^{-1}(\omega)&&\mathbf{h}_k^{-1}(-\omega)\\
\mathbf{h}_k^{-1}(\omega)&&0
\end{pmatrix}
\end{equation}
Thus the flow equation of the equilibrium function $F_{k}[\varphi]$, now writes:
\begin{align}\label{modred}
    \partial_{k}F_{k}[\varphi; x] & = \frac{1}{2}\tilde{\partial_{k}}\int_{t,u, v}\Tr\left\{(\Gamma^{(2)}_{k}+R_{k})^{-1}.\frac{\delta\Gamma^{(2)}_{k}}{\delta\tilde{\varphi}}\right\}|_{\tilde{\varphi}=0, \varphi =\varphi(x)} \nonumber\\
& = \frac{1}{2}\tilde{\partial_{k}}\Tr\int_{\omega}\int_{u, v}\sum^{2n}_{p\geq 1}\mathbf{h}^{-1}_{k}(-\omega)\cdot A^{(p)}_{k}(\varphi)(-i\omega)^{p-1}+\mathbf{h}^{-1}_{k}(\omega)\cdot A^{(p)}_{k}(\varphi)(i\omega)^{p-1}\nonumber\\
& +\frac{1}{2}\tilde{\partial_{k}}\Tr\int_{\omega}\int_{u, v}\frac{1}{i\omega}\left(\mathbf{h}_k^{-1}(\omega)-\mathbf{h}_k^{-1}(-\omega)\right)\cdot\left(F''_{k}[\varphi; x]+\sum^{n}_{p\geq 1}D^{(p)}_{k}(\varphi)\omega^{2p}\right)\nonumber\\
&-\frac{1}{2}\tilde{\partial_{k}}\Tr\int_{\omega}\int_{u, v}\left[\left\{\mathbf{h}_k^{-1}(-\omega)\cdot\left(\sum_{p\geq 1}\partial_{\varphi}\left(M^{(p)}_{k}(\varphi)\varphi\right)(1+(-1)^p)(i\omega)^{p-1}\right)\cdot \mathbf{h}_k^{-1}(\omega)\right\}\cdot\left(F''_{k}[\varphi; x]+\sum^{n}_{p\geq 1}D^{(p)}_{k}(\varphi)\omega^{2p}\right)\right].
\end{align}
The flow of the equilibrium function $F_{k}[\varphi; x]$ is thus modified by the term  in the last line of Eq.~\eqref{modred}. One must notice that this term is actually odd-in $\omega$. Thus as long the $\omega-$integral is convergent, this term always vanishes. This is indeed the case (after acting with the $\tilde{\partial}_{k}$), and hence the last term does not contribute to the flow of $F_{k}[\varphi; x]$ and we are left with Eq.\eqref{OmegaL}. 
\bibliography{main}
\bibliographystyle{apsrev4-1}
\end{document}